\documentclass{aa}  
\usepackage{graphicx}
\usepackage{txfonts}
\begin{document}
   \title{SN 1999ga: a low-luminosity linear type II supernova?}

   \subtitle{}

   \author{A. Pastorello\inst{1}
	  \and
	  R. M. Crockett\inst{1} 
          \and
          R. Martin\inst{2}
          \and
	  S. J. Smartt\inst{1}
	  \and
	  G. Altavilla\inst{3}
	  \and
          S. Benetti\inst{4}
	  \and
	  M. T. Botticella\inst{1}
	  \and 
	  E. Cappellaro\inst{4}
	  \and
	  S. Mattila\inst{5}
	  \and	
	  J. R. Maund\inst{6,7}
	  \and
          S. D. Ryder\inst{8}
	  \and	      
	  M. Salvo\inst{9}
	  \and
	  S. Taubenberger\inst{10}
	  \and
	  M. Turatto\inst{11}
}

   \offprints{A. Pastorello}

   \institute{Astrophysics Research Centre, School of Mathematics and Physics, Queen's University
 Belfast, Belfast BT7 1NN, United Kingdom\\
              \email{a.pastorello@qub.ac.uk}
         \and
	 Perth Observatory, 337 Walnut Road, Bickley 6076, Perth, Australia 
         \and
	 INAF - Osservatorio Astronomico di Bologna, Via Ranzani 1, I-40127, Bologna, Italy
         \and
	 INAF - Osservatorio Astronomico di Padova, Vicolo dell' Osservatorio 5, I-35122 Padova, Italy
         \and
	 Tuorla Observatory, Department of Physics \& Astronomy, University of Turku, V\"ais\"al\"antie 20, FI-21500, Piikki\"o, Finland
         \and
         Dark Cosmology Centre, Niels Bohr Institute, University of Copenhagen, Julian Maries Vej 30, 2100 Copenhagen, Denmark
	 \and
	 Sopie and Tycho Brahe Fellow
         \and
	 Anglo-Australian Observatory, PO Box 296, Epping, NSW 1710, Australia
         \and
	 Research School of Astronomy and Astrophysics, Australian National University, Mount Stromlo and\\ Siding Spring Observatories
	 Cotter Road, Weston Creek, ACT 2611, Australia
         \and
	 Max-Planck-Institut f\"ur Astrophysik, Karl-Schwarzschild-Str. 1, D-85741 Garching bei M\"unchen, Germany
         \and
	 INAF - Osservatorio Astrofisico di Catania, Via S. Sofia 78, I-95123 Catania, Italy
}
  

   \date{Received  XXXXX ; accepted XXXXX}

 
  \abstract
   {Type II-linear supernovae are thought to arise from progenitors that have lost most of
   their H envelope by the time of the explosion, and they are poorly understood because they are only
   occasionally discovered. It is possible that they are intrinsically
   rare, but selection effects due to their rapid luminosity evolution may also play an important role 
   in limiting the number of detections. In this context, the discovery of a subluminous type II-linear 
   event is even more interesting.}
   {We investigate the physical properties and characterise the explosion site of the type II SN 1999ga, which exploded 
   in the nearby spiral galaxy NGC 2442.}
   {Spectroscopic and photometric observations of SN 1999ga allow us to constrain the energetics of the explosion and
   to estimate the mass of the ejected material, shedding light on the nature of the progenitor star in the final stages of
   its life. The study of the environment in the vicinity of the 
   explosion site provides information on a possible relation between these unusual supernovae and 
   the properties of the galaxies hosting them.}
   {Despite the lack of early-time observations, we provide reasonable evidence that SN 1999ga was probably 
   a type II-linear supernova that
   ejected a few solar masses of material, with a very small amount of radioactive elements of the order of 0.01M$_\odot$.}
   {}

   \keywords{stars: supernovae: general --
                stars: supernovae: individual: SN 1999ga --
                stars: supernovae: individual: SN 1979C --
		stars: supernovae: individual: SN 1980K --
		stars: supernovae: individual: SN 1990K
               }

   \maketitle
%

\section{Introduction}

Type II-linear supernovae (SNe IIL) form a rare and poorly studied class of core-collapse supernovae (SNe). 
Contrary to type II-plateau supernovae 
(SNe IIP) that show slow photometric evolution 
and well-correlated observed properties (see e.g. \cite{ham03} and references therein), SNe IIL  are 
characterised by fast-evolving light curves and a higher degree of heterogeneity. 
SNe IIL are believed to arise either from the explosion of moderate-mass
progenitors (8-10M$_\odot$) that have lost a significant 
fraction of their H envelope via binary interaction or from 
more massive stars that lose mass before their explosion through strong stellar winds.  

However, despite the mass loss, a significant amount of H still remains in the stellar envelope,
and, consequently, the spectrum of a SN IIL still shows prominent H lines.
The evidence that in SNe IIL the H envelope is not as massive as in SNe IIP comes mostly from 
analysis of the light curves. Because of the lack of 
a massive H envelope that recombines with cooling, the light curves do not show 
the characteristic long period (about three months) of almost constant luminosity typical of 
SNe IIP, but instead show a linear decline after maximum.

In Patat et al. (1993) and Patat et al. (1994) a compilation of light curves from the literature of a large number of SNe II was analysed. 
A high degree of heterogeneity in the light curve shapes was found, and many transitional objects 
between type IIP and  IIL SNe were shown to exist. This is probably a consequence of the progressively declining mass 
of the residual H envelope along the sequence
from type IIP to type IIL events. Up to the point of explosion, the progenitors of type IIP SNe
have retained much of their H-rich envelopes.
On the other hand, stars generating SNe IIL have already lost 
most of their H envelopes at the time of the explosion. 
In a few cases, the presence of this  circumstellar material (CSM)
is revealed at some point because of the interaction with the expanding ejecta. However, in general, this
is not the dominant contribution to the luminosity of SNe IIL, at least at early phases.
A consequence of this is that SNe IIL show a much more rapid photometric evolution than type IIn SNe (\cite{sch90}), 
in which the ejecta-CSM interaction is the main process regulating the output of energy of the SN even at early phases. 
Nevertheless, a few type II SNe show unequivocal evidence of interaction with CSM  from specific spectral properties 
and the high peak luminosity without showing slowly evolving light curves.
This is the case for a few SNe IIn with fast-declining light curves, 
like SN 1998S (e.g. \cite{fas00,liu00}), SN 1999el (\cite{elisa02}), or SN 2006jc (and SNe Ibn) which interact with a He-rich CSM 
(\cite{pasto07a,foley07,pasto08a}). All these SNe have light curves with rapid evolution resembling those of SNe IIL.
    
The prototype of the classical SNe IIL is SN 1979C, which exploded in M100 and is by far the best
studied object of this class, being extensively observed from X-ray to radio wavelengths 
(see e.g. \cite{pan80,deva81,bra81,wei91,fes93,imm98,fes99,van99,mon00,ray01,mar02,bar03,imm05a}). 
Other well studied SNe IIL discovered in the last 30 years are 
SN 1980K in NGC6946 (see e.g. \cite{buta82,bar82} for early-time data, and \cite{uom86,fes99}, and 
references therein, for late-time data),
SN 1986E in NGC 4302 (\cite{cap90,cap95a}) and SN 1990K in NGC 150 (\cite{cap95b}).
 
A few SNe IIL, sometimes dubbed as type IId SNe (\cite{ben00}), show narrow P-Cygni Balmer lines of hydrogen
superimposed on otherwise normal broad-lined spectra of type II SNe. This is interpreted as
an evidence of super-winds occurred years to decades before the SN explosions. 
This group includes SNe 1994aj, 1996L, 1996al, 2000P
(\cite{ben98,ben99,ben96,ben00,cap00,jha00}). 
In all these cases, ejecta-CSM interaction takes place a few years after the explosion,
when the fast SN ejecta reaches the slower wind.
The ejecta-CSM interaction may be heralded by a number of 
different pieces of observational evidence, e.g. flattening of the late-time optical light curves, the peculiar
profile of spectral lines (e.g. \cite{sch90,tura93,lei91,fes99}), strong X-ray (see \cite{sch95,imm98,asc07} for reviews)
and radio emission (e.g. \cite{van96}).

A special mention has to be given to the recently discovered type IIL SNe that have extremely 
high luminosities (M$_V <$  -22.5). This small sub-group includes SN 2005ap (\cite{qui07}) and SN 2008es (\cite{mil08,gez08}).
Their extraordinary luminosity and their spectra, which show no
evidence of the narrow features typical of interacting objects (SNe IIn), are
best explained by strong interaction of the ejecta with a dense, optically-thick CSM, rather than
with very large ejected $^{56}$Ni masses (\cite{qui07}).

In this paper we study the case of SN 1999ga, a type II SN (likely of type IIL) which was followed only at late
phases. Some pre-explosion, ground-based images, and very late time Hubble Space Telescope (HST) frames have been studied 
in order to constrain the characteristics of the progenitor star and the site of explosion.
Basic information on SN~1999ga and its spectacular host galaxy is reported in Sect. \ref{ga}.
Photometric and spectroscopic data are presented in Sect. \ref{lc} and Sect. \ref{sp}, respectively. 
The explosion site, imaged in pre-SN archival observations and in deep HST frames obtained 
years after the SN explosion, are analysed in Sect. \ref{hb}, while a discussion and a summary follow
in Sect. \ref{dc}.

   \begin{figure}
   \centering
   \includegraphics[angle=0,width=9cm]{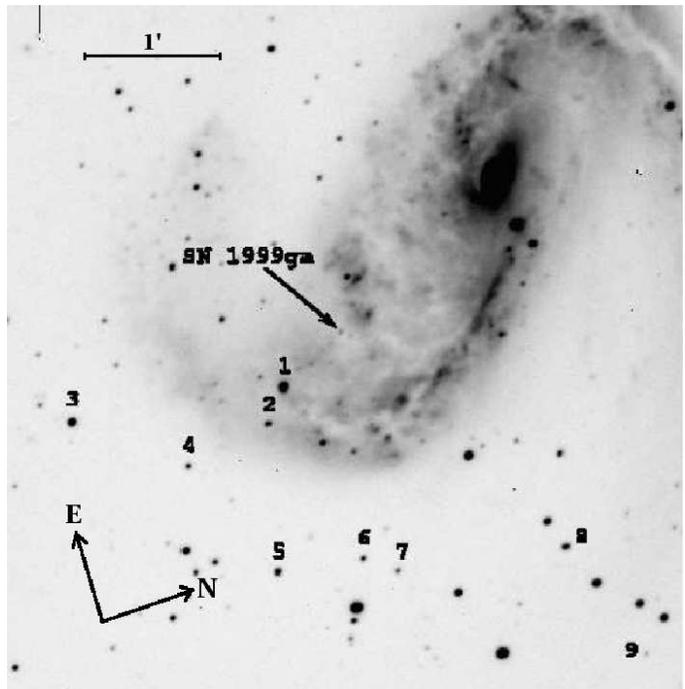}
   \caption{SN 1999ga and NGC 2442. R-band image obtained on Apr 7, 2000 with the ESO 3.6m Telescope equipped with EFOSC2. A 
   sequence of reference stars is marked by numbers. }
              \label{Fig1}
    \end{figure}

\section{SN 1999ga in NGC 2442} \label{ga}

SN 1999ga was discovered on November 19.76 UT in the spiral galaxy NGC 2442 by \cite{woo99}, 
on behalf of the Perth Astronomy Research Group (PARG, see e.g. \cite{woo98}),
in the course of an automated supernova search performed with the Perth-Lowell 0.61-m reflector. 
The discovery magnitude quoted by Woodings et al. (1999), R$\sim$18, is wrong, being almost 2 mag fainter than the true 
magnitude of SN~1999ga at that epoch (cf. Sect. \ref{lc}). The exact position was measured
on Nov. 22.54 with the 0.25-m Mike Candy Telescope at Perth Observatory and found
to be R.A. = $7^{h}36^{m}16\fs70 \pm 0\fs03$, Dec. = $-69\degr33\arcmin21\farcs8 \pm 0\farcs4$ (equinox
2000.0), about 38$\arcsec$ west and 91$\arcsec$ south of the centre of NGC 2442 (\cite{woo99}).

\begin{table}
\caption{Magnitudes of reference stars in the SN field. The r.m.s. of the average magnitudes
are reported in brackets.}\label{Tab1}
\centering
\begin{tabular}{ccccc}
\hline \hline
Star & B & V & R & I  \\ \hline
1 & 17.55 (0.02) & 16.55 (0.01) & 15.89 (0.02) & 15.18 (0.02) \\ 
2 & 20.02 (0.02) & 18.48 (0.01) & 17.52 (0.01) & 16.37 (0.02) \\
3 & 18.29 (0.03) & 16.84 (0.01) & 15.92 (0.05) & 15.05 (0.03) \\
4 & 19.45 (0.02) & 18.73 (0.01) & 18.23 (0.01) & 17.73 (0.02) \\ 
5 & 18.82 (0.01) & 18.05 (0.01) & 17.59 (0.01) & 17.09 (0.01) \\
6 & 20.33 (0.02) & 18.76 (0.01) & 17.77 (0.01) & 16.71 (0.02) \\
7 & 20.03 (0.01) & 19.15 (0.02) & 18.57 (0.01) & 18.08 (0.02) \\
8 & 19.41 (0.04) & 18.21 (0.03) & 17.38 (0.04) & 16.80 (0.03) \\
9 &              & 20.36 (0.04) &              & 18.05 (0.02) \\
\hline
\end{tabular}
\end{table}

   \begin{figure}
   \centering
   \includegraphics[angle=0,width=8.8cm]{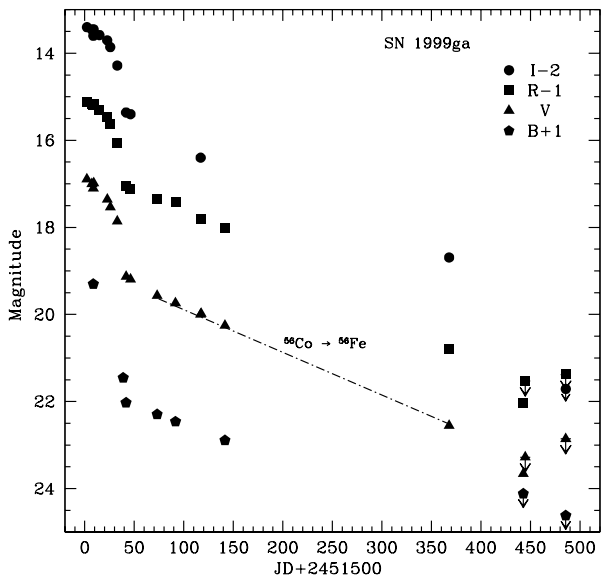}
   \caption{B, V, R, I light curves of SN 1999ga. Unfortunately, only late time photometry is available for this
   SN, because the SN was not observed at early epochs as the host galaxy was too low on the horizon for most of the
   night. 
   Another seasonal gap is visible between JD=2451650 and JD=2451850.
   The photometric points from \protect\cite{rub99} are also included.}
              \label{Fig2}
    \end{figure}

\begin{table*}
\caption{Calibrated magnitudes of SN 1999ga. An additional detection limit in
the U band (U $>$ 21.9) was obtained on 1999, December 29.}\label{Tab2}
\centering
\begin{tabular}{cccccccc}
\hline \hline
Date & JD & B & V & R & I & Instrument \\ \hline
1999-11-19 & 2451502.29 &              & 16.90 (0.08) & 16.11 (0.05) & 15.40 (0.04) & PARG \\
1999-11-24 & 2451507.29 &              & 17.00 (0.07) & 16.18 (0.06) & 15.46 (0.06) & PARG \\
1999-11-26 & 2451509.02 &              & 16.98 (0.04) & 16.18 (0.03) & 15.45 (0.04) & PARG \\
1999-12-02 & 2451514.84 &              &              & 16.30 (0.03) & 15.58 (0.05) & PARG \\
1999-12-10 & 2451522.79 &              & 17.36 (0.25) & 16.46 (0.04) & 15.70 (0.14) & PARG \\
1999-12-13 & 2451525.84 &              & 17.53 (0.05) & 16.62 (0.04) & 15.86 (0.04) & PARG \\ 
1999-12-20 & 2451532.87 &              & 17.86 (0.07) & 17.05 (0.05) & 16.28 (0.07) & PARG \\
1999-12-26 & 2451538.82 & 20.45 (0.07) &              &              &              & WFI$^\star$ \\ 
1999-12-29 & 2451541.77 & 21.03 (0.08) & 19.13 (0.03) & 18.05 (0.04) & 17.36 (0.06) & DF \\
2000-01-01 & 2451546.27 &              & 19.19 (0.42) & 18.11 (0.21) & 17.40 (0.32) & PARG $^\diamond$ \\
2000-01-29 & 2451573.13 & 21.29 (0.30) & 19.57 (0.04) & 18.35 (0.07) &              & AAT $^\ast$\\   
2000-02-17 & 2451591.67 & 21.46 (0.13) & 19.74 (0.07) & 18.42 (0.03) &              & WFI$^\star$ \\ 
2000-03-13 & 2451617.02 &              & 20.0   (0.3) & 18.8   (0.3) & 18.4  (0.4)  & DF$^\dag$ \\
2000-03-14 & 2451617.5  &              & 19.98 (0.07) &              &              & DF \\   
2000-04-07 & 2451641.57 & 21.89(0.05)  & 20.26 (0.03) & 19.01 (0.07) &              & EF2 \\ 
2000-11-19 & 2451867.85 &              & 22.55 (0.24) & 21.79 (0.21) & 20.69 (0.24) & DF \\
2001-02-02 & 2451942.75 & $>$23.1      & 23.66 (0.38) & 23.02 (0.47) &              & EF2 \\
2001-02-04 & 2451944.63 &              & $>$ 23.3     & $>$22.5      &              & WFI \\
2001-03-17 & 2451985.61 & $>$23.6      & $>$22.9      & $>$22.4      & $>$23.7      & DF \\
\hline
\end{tabular}

$^\ast$ Anglo Australian Telescope Data Archive: {\sl
http://archive.ast.cam.ac.uk/arc-bin/wdb/aat$\_$database/observation$\_$log/make};\\
$^\star$ ESO Data Archive: {\sl http://archive.eso.org/eso/eso$\_$archive$\_$main.html};
$^\diamond$ poor night;
$^\dag$ spectrophotometric magnitude \\

PARG = Perth-Lowell 0.61-m Cassegrain Telescope + CCD (Perth, Australia);
EF2 = ESO 3.6m Telescope + EFOSC2 (La Silla, Chile);\\
WFI = ESO/MPI 2.2m Telescope + Wide Field Imager (La Silla, Chile);
DF = Danish 1.54m Telescope + DFOSC (La Silla, Chile).\\

\end{table*}

Additional observations, obtained on November 25.18 UT by Rubinstein (1999)
at the Cerro Tololo Interamerican Observatory, indicated that the magnitude of
the transient was significantly brighter than that reported by Woodings et al. (1999)
(B=18.3$\pm$0.3, V=17.1$\pm$0.1, R=16.2$\pm$0.1, V=15.6$\pm$0.1).
On the basis of the SN location alone, Rubinstein (1999) suggested a possible
type II classification for SN~1999ga. Because of its red colour, 
he also suggested significant interstellar reddening.
However, we will see in Sect. \ref{lc} and Sect. \ref{sp} that SN 1999ga
was discovered late, and it is well known that type II SNe are intrinsically red 
at the end of the photospheric phase (e.g. \cite{pasto04}).

The tentative SN classification of Rubinstein (1999) was confirmed spectroscopically
by Salvo et al. (1999) on the basis of a spectrum taken on December 29 at ESO-La Silla. 
The classification spectrum (presented in this paper) shows
that SN~1999ga is a type II SN  observed a few months after explosion. 
In this paper we will adopt JD = 2451420 as an indicative epoch for the explosion of SN 1999ga, 
which was derived via comparisons with observed data of well-studied type II SNe (Sect. \ref{lc} and Sect. \ref{sp}).

The spectral lines in the classification spectrum show  P-Cygni profiles, while H$\alpha$ has a peculiar
 flat-topped emission component  indicative that the H$\alpha$ emission mostly comes from a shell-like
region (see Sect. \ref{sp} for discussion).
Finally the spectrum shows evidence of little 
interstellar extinction (\cite{sal99}), suggesting moderate reddening.

The spectacular galaxy hosting SN 1999ga, NGC 2442 (Fig. \ref{Fig1}), is classified by  
HyperLeda\footnote{http://leda.univ-lyon1.fr/} 
as an SBbc galaxy, with well-developed but asymmetric spiral arms. It belong to the Volans group 
of galaxies, and Ryder et al. (2001) report a distance of about 15.5 Mpc (distance modulus $\mu$ = 30.95 mag). 
From the recessional velocity corrected by Local Group infall into Virgo, 1150 km s$^{-1}$, and assuming H$_0$ = 72 km s$^{-1}$
Mpc$^{-1}$, we obtain a 
slightly higher distance, d $\approx$ 16 Mpc ($\mu$ = 31.02 mag), which
will be adopted throughout this paper. The Galaxy extinction in the direction of NGC 2442 is
rather high, being E(B-V)$_{Gal}$ = 0.20 mag (\cite{sch98}). The low signal to noise and the late phase of the 
SN 1999ga spectra (see Sect. \ref{sp}) do not allow precise measurement of the equivalent width (EW) of Galaxy 
and host galaxy interstellar Na I doublet (Na ID) absorptions. However, we tentatively estimated the two components
of Na ID in the SN spectra and found that the NGC 2442 component  has an EW $\approx$ 0.9\AA,
which is roughly 70 per cent that of the Galaxy. This would imply an E(B-V)$_{host}$ = 0.14 mag, which  
is consistent with what we find using the correlation between EW of the interstellar Na ID and E(B-V) of Turatto et al.
(2003). 
Accounting for both the Galactic reddening component and that of the host galaxy, we obtain 
E(B-V)$_{tot}$ = 0.34 mag as our best estimate for the total reddening, and this value will be adopted hereafter.


Several attempts have been made in order to explain the 
disturbed morphology of NGC 2442, by searching for evidence of tidal interaction with a few nearby galaxies (\cite{elm91,san94,mih97,hou98}).
\cite{ryd01} found evidence that NGC 2442 is associated with an extremely massive cloud of H I, with a
mass almost one third that
of the galaxy itself. This gas cloud was probably produced in a recent tidal encounter with a moderately massive companion, 
though both ram pressure-stripping, and HI
rings/arcs stripped from the outer envelope of a low surface brightness galaxy (\cite{bek05})
are also possible scenarios. Moreover, Bajaja et al. (1995), Mihos $\&$ Bothun (1997) and Bajaja et al. (1999) found an elliptical, circumnuclear, 
star-forming molecular ring. Finally, the intensity ratios of emission lines of the galaxy nucleus
indicate that NGC 2442 is likely a LINER (\cite{baj99}).

\section{Light Curve} \label{lc}

Follow-up photometric observations were carried out using a number of telescopes in Australia and Chile.
Available imaging of SN~1999ga was reduced following a standard procedure (see e.g. \cite{pasto05}):
the images were first overscan, bias and flat-field 
corrected. SN magnitudes were then measured using a PSF-fitting technique after the subtraction of template images. The recovered 
magnitudes were then scaled using the night zero-points 
computed comparing the instrumental magnitudes of several well-known standards fields with those reported in the Landolt
catalogue (\cite{lan92}). Finally the SN magnitudes were fine tuned with reference to the magnitudes of 7 stars in the field of NGC 2442 
(see Fig. \ref{Fig1}) obtained by averaging the estimates obtained in photometric nights. The magnitudes of the sequence 
stars, as denoted in Fig. \ref{Fig1}, are reported in Tab. \ref{Tab1}, while the B, V, R and I magnitudes of SN~1999ga~are
given in Tab. \ref{Tab2}.

   \begin{figure}
   \centering
   \includegraphics[angle=0,width=9cm]{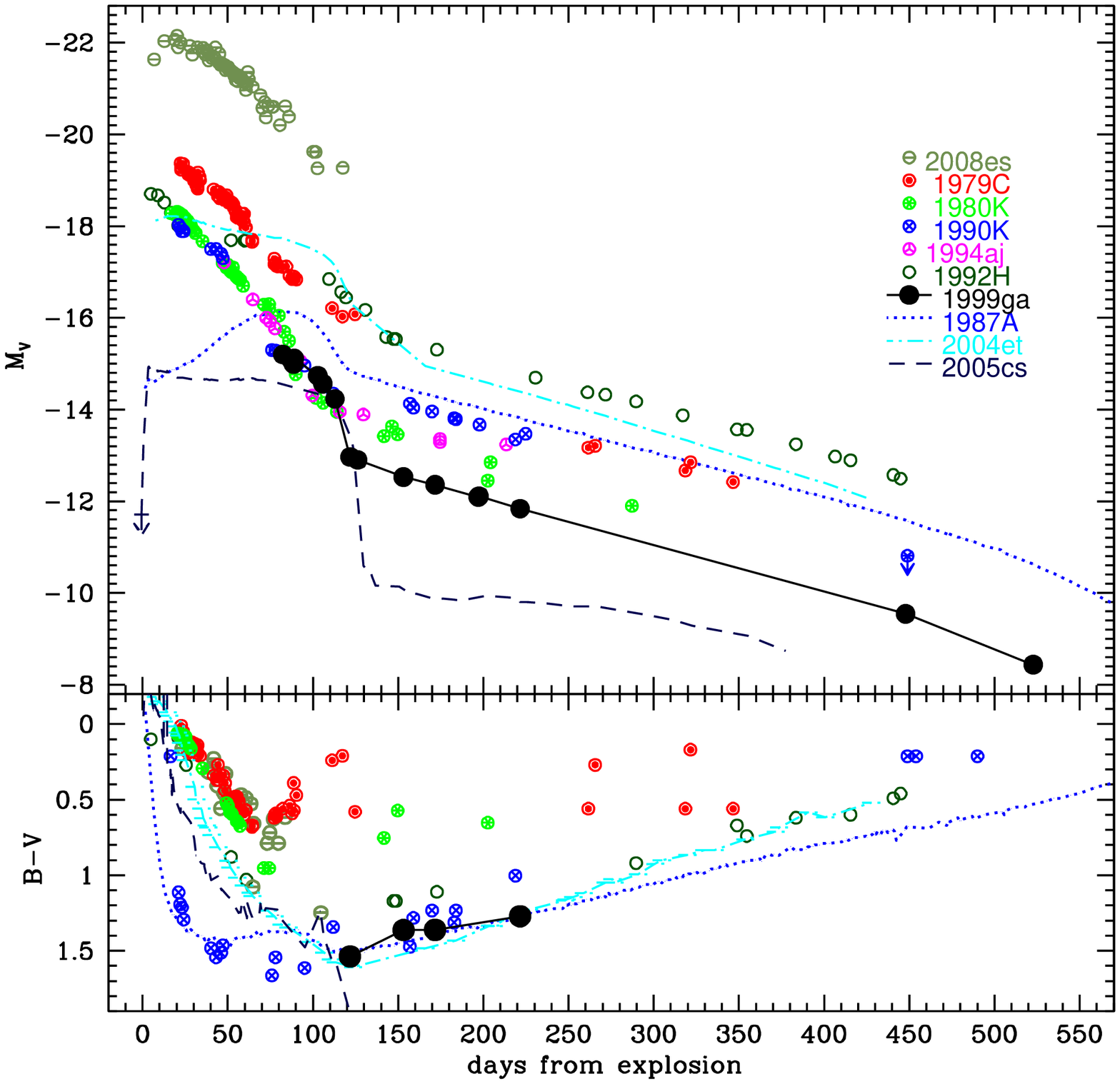}
   \caption{Top: V-band absolute light curves of SN 1999ga and a number of type II SNe: the high-luminosity SN 2008es (\cite{gez08,mil08}), 
   the type IIL SNe 1979C (\cite{bal80,deva81,bar82b}), 1980K (\cite{bar82,buta82,tsv83}), 1990K (\cite{cap95b}), 
   1994aj (\cite{ben98}); the transitional type IIP/IIL SN 1992H (\cite{clo96}); the peculiar SN 1987A (\cite{whit89} and references therein) and the type IIP SNe 2004et
   (\cite{sahu07,kun07}) and 2005cs (\cite{tsv06,pasto06,pasto09}). The explosion epochs of SNe IIL have been estimated 
   to occur roughly 2-3 weeks before their maximum light (\cite{pat93}).
   Bottom: comparison of the B-V colour curves of SNe 1999ga, 2008es, 1979C, 1980K, 1990K, 1992H, 1987A, 2004et.}
              \label{Fig3}
    \end{figure}

In Fig. \ref{Fig2} the photometric evolution of SN~1999ga in the optical bands is shown.
The late discovery of SN~1999ga prevents us from studying the early evolution of its light curve,
with particular reference to the immediate post-maximum behaviour. As a consequence, 
uncertainty as to the shape of its light curve (linear or plateau) remains, although there 
is some evidence in favour of a type IIL classification for SN 1999ga (see Sect. \ref{sp}).

SN~1999ga was discovered during the post-maximum decline phase. The light curve shows a rapid luminosity drop (by 1.5-2 mag in all bands), followed by a less steep decline
corresponding to the exponential (radioactive) tail. As evidenced in Fig. \ref{Fig2}, 
at late time the V-band light curve of SN~1999ga closely matches
the $^{56}$Co $\rightarrow ^{56}$Fe decay slope (0.98 mag/100$^d$, assuming complete gamma-ray trapping), even though at very late phases (from $\sim$400 days after explosion) 
the light curves in all bands show steeper declines. This is has been seen in other type II SNe, and it 
is generally interpreted as dust forming in the SN ejecta and/or incomplete $\gamma$-ray trapping (e.g. \cite{abou03}).

The slopes in the period between 120 and 450 days, as derived from a least squares fit to the light 
curves, are $\gamma_B$=0.87 mag/100$^d$, $\gamma_V$=1.02 mag/100$^d$, $\gamma_R$=1.15 mag/100$^d$ and 
$\gamma_I$= 1.00 mag/100$^d$, on average rather consistent with those expected by the radioactive 
decay of $^{56}$Co in the case of complete $\gamma$-ray trapping. This 
might be an indication of relatively massive ejecta.

In Fig. \ref{Fig3} (top panel) the V-band absolute light curve of SN 1999ga is compared with
those of a number of type II SNe, four SNe IIL (1979C, 1980K, 1990K and the peculiar 1994aj, which belongs to the IId
sub-type, \cite{ben00}), the over-luminous type IIL SN 2008es, the transitional type IIP/IIL SN 1992H  and three SNe IIP (2004et, 2005cs and the peculiar 1987A). 
The phases for SN 1999ga have been computed assuming JD = 2451420 as an indicative epoch for the core-collapse,
as derived from a comparison of the light curve and spectra with those of other SNe II.
The apparent flattening in the very early
light curve of SN 1999ga would suggest a plateau-like behaviour (or, at least, a transitional object). However, Patat et al. (1993, 1994) 
noted that a shoulder in 
the light curves of SNe IIL (especially in the R band) is frequently observed after maximum. Hence the available photometry 
of SN 1999ga does not allow us to definitely discriminate between a {\sl linear} or a {\sl plateau} type light curve. We will see in Sect. \ref{sp}
that more clear clues in favour of a type IIL classification for SN 1999ga come from the spectroscopy.

\begin{table*}
\caption{Spectroscopic observations of SN~1999ga. The spectrum at 39.5 days is the one used to classify SN 1999ga
(\protect\cite{sal99}).}\label{Tab3}
\centering
\begin{tabular}{ccccccc}
\hline \hline
Date & JD(+2400000) & Phase$^\dag$ & Instrumental configuration & Exposure & Resolution (\AA) & Range (\AA) \\ \hline
1999-12-29 & 51541.76 & 39.5 & Danish 1.54m + DFOSC + gr.4 & 1200s & 11 & 3650-9050 \\
2000-03-12/13 & 51617.02$^\ddag$ & 114.8 & Danish 1.54m + DFOSC + gr.4 + gr.5 & 1800s + 1800s & 11+12 & 3570-10200 \\
2000-04-07 & 51641.58 & 139.3 & ESO 3.6m + EFOSC2 + gr.11 & 2 $\times$ 1800s & 18 & 3380-7500 \\
2001-02-02 & 51942.71 & 440.5 & ESO 3.6m + EFOSC2 + gr.11 & 3 $\times$ 1800s & 18 & 3490-7460 \\
\hline
\end{tabular}

$^\dag$ Days from discovery (JD=2451502.26);
$^\ddag$ Spectrum obtained averaging two spectra obtained in subsequent days.
\end{table*}

With the assumption on the explosion epoch mentioned above, 
a comparison between the integrated late-time BVRI luminosity of SN 1999ga with that of SN 1987A at similar epochs provides an approximate 
estimate of the $^{56}$Ni mass synthesized by SN 1999ga, being about 0.013 $\pm$ 0.003 M$_\odot$, which is to our knowledge the smallest ever 
registered for a type IIL SN, and only marginally higher than those registered for the low-luminosity 
SNe IIP (e.g. \cite{pasto04}). 

A comparison among B-V colour curves of SN 1999ga and some other SNe II is also shown in Fig. \ref{Fig3} (bottom panel).
Surprisingly, the overall behaviour of the colour curve of the type IIL SN 1990K within the first $\sim$250 days closely resembles that of SN 1987A, more than 
that of other SNe IIL shown in Figure. However, we should remember that the luminous light curve peaks and the bluer colours of 
SN 1979C and SN 1980K are probably signatures of interaction with a CSM (see Sect. \ref{dc}).  
Unfortunately, only a few points are available for SN 1999ga and this does not allow us to study in detail its colour 
evolution. At $\sim$120 days, the B-V colour of 1999ga is around 1.5 mag, then it becomes marginally bluer (by about 1.3 mag) 4 months
later. However, the overall similarity with the colour evolutions of SNe 1987A, 2004et, 1992H and 1990K
 (between +110 and +230 days) is an additional argument in favour of a late discovery of SN 1999ga.

\section{Late Spectral Evolution} \label{sp}

   \begin{figure}
   \centering
   \includegraphics[angle=0,width=9.3cm]{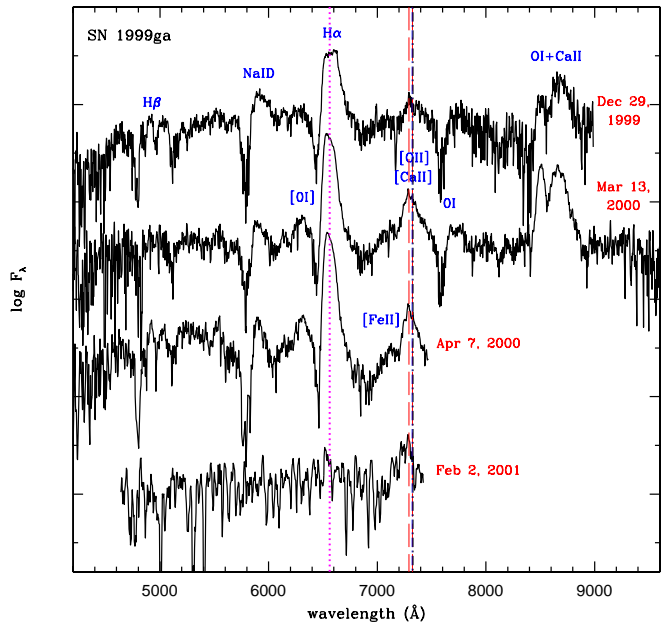}
   \caption{Sequence of nebular spectra of SN~1999ga. The main features are labelled. The vertical lines mark the rest
   positions 
   of H$\alpha$ (magenta, dotted line), [Ca II] $\lambda\lambda$7291,7324 (red, dashed lines) and [O II] $\lambda\lambda$7320,7330 (blue, dot-dashed lines). There is minor evidence of residual 
   H$\alpha$ in the February 2, 2001 spectrum. All spectra, reported at the host galaxy rest wavelength, have been reddening corrected.}
    \label{Fig4}
    \end{figure}

   \begin{figure}
   \centering
   \includegraphics[angle=0,width=9cm]{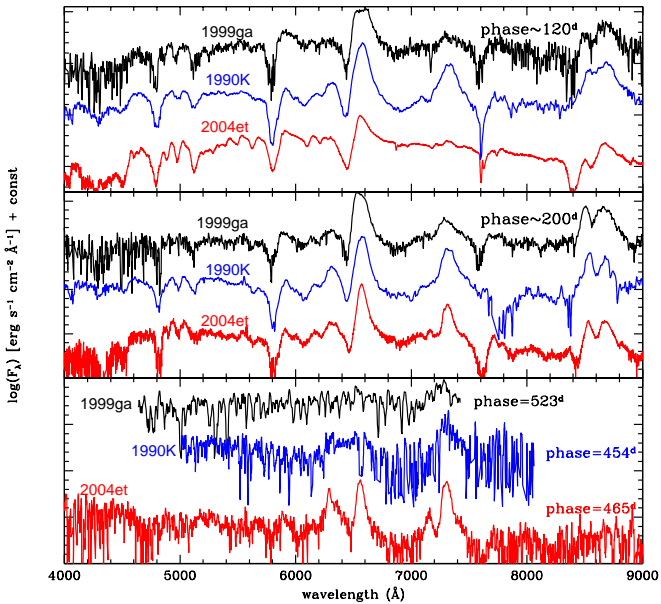}
   \caption{Comparison between spectra of SN 1999ga, the type IIL SN 1990K (\cite{cap95b}) and the type IIP SN 2004et (\cite{sahu07}) 
   at different phases. The phases are computed with reference to an approximate estimate of the explosion dates. For SN 1990K we adopted 
   JD = 2448020 as explosion time, which is about 17 days
   before the discovery epoch (JD=2448037.3). Our assumption is in good agreement with that of Cappellaro et al. (1995b), who estimated 
   the explosion of SN 1990K to occur about two weeks before its discovery. Finally, for SN 2004et we adopt the same explosion 
   epoch as Sahu et al. (2006) (JD = 2453270.5, see also \cite{li05}).}
              \label{Fig5}
    \end{figure}

\begin{table*}
\caption{Pre- and late, post-explosion (ground-based and HST) images. The magnitudes of {\it source A} in the
ground-based observations are also reported in column 5.}\label{Tab4}
\centering
\begin{tabular}{ccccccc}\hline \hline
Date & JD & Telescope & Filter & Exposure time (s) & {\it Source A} magnitude & Proposal ID / Source \\ \hline
1989-12-31 & 2447891.53 & SSO1m & H$\alpha$ & 3$\times$1000 & & \protect\cite{ryd93}\\
1989-12-31 & 2447891.57 & SSO1m & R$_c$ & 2$\times$500 & $>$18.76 & \protect\cite{ryd93}\\
1990-03-04 & 2447955.45 & SSO1m & V & 2$\times$250 & $>$20.70 & Obs. S. Ryder \\
1991-01-11 & 2448267.62 & SSO1m & I & 2$\times$250 & $>$20.20& Obs. S. Ryder \\
1995-02-22 & 2449771.06 & AAT     & unfilt.& 300+60 & 21.85$\pm$0.40$^\ddag$ & Obs. Whiteoak $\&$ Koribalski \\
1995-03-01 & 2449777.57 & CTIO1.5m & B & 2$\times$600 & 22.76$\pm$0.32 & Obs. G. Purcell \\
1995-03-01 & 2449777.58 & CTIO1.5m & I & 300 & 21.37$\pm$0.37 &Obs. G. Purcell \\ \hline
2006-01-28 & 2453764.70 & ESO2.2m & H$\alpha$ & 4$\times$720 & & 076.C-0888 (PI: Y. Bialetski)\\
2006-01-29 & 2453764.77 & ESO2.2m & V & 2$\times$600  & 22.07 $\pm$ 0.09 & 076.C-0888 (PI: Y. Bialetski) \\
2006-01-30 & 2453765.73 & ESO2.2m & B & 2$\times$300  & 22.89 $\pm$ 0.08 & 076.C-0888 (PI: Y. Bialetski) \\
2006-11-29 & 2454068.78 & ESO2.2m & B & 4$\times$600  & 22.87 $\pm$ 0.22 & MPI Time (PI: W. Hillebrandt) \\ \hline
2006-10-21 & 54030.31 & HST & F435W & 1580 & & 10803 (PI: S. Smartt) \\
2006-10-21 & 54030.37 & HST & F658N & 1350 & & 10803 (PI: S. Smartt)  \\
2006-10-21 & 54030.39 & HST & F814W & 1200 & &  10803 (PI: S. Smartt) \\ \hline
\end{tabular}

$^\ddag$ magnitude from unfiltered image, rescaled to the R band photometry.\\
SSO1m =  1m-Telescope + CCD (Siding Spring Observatory, Australia); \\
AAT = 4m Anglo Australian Telescope + CCD (Siding Spring Observatory, Australia); \\
CTIO1.5m = 1.5m-Telescope + CCD (Cerro Tololo Inter-American Observatory, Chile)\\
ESO2.2m = ESO/MPI 2.2m Telescope + Wide Field Imager (La Silla, Chile); \\
HST = Hubble Space Telescope + ACS/WFC Camera.\\
\end{table*}

Four spectra of SN 1999ga have been obtained with the ESO telescopes at La Silla (Chile), and 
basic information about these spectra is reported in Tab. \ref{Tab3}. The available spectral sequence 
is shown in Fig. \ref{Fig4}.

The earliest spectrum is typical for a type II SN transiting from the photospheric to the 
nebular phase. Prominent H$\alpha$ and H$\beta$ are visible, with relatively weak P-Cygni absorptions, along with 
strong P-Cygni lines of Na ID and O I $\lambda$7774. 
The minima of the absorption 
components of H$\alpha$ and Na ID are blue-shifted by about 5800 km s$^{-1}$ and 5300 km s$^{-1}$, respectively.    
The feature at about 8600\AA~ is
attributed to a blend of O I $\lambda$8446 plus the Ca II near-infrared triplet. 
A few Fe II lines are possibly detected near the region of H$\beta$. 
It is worth to note that H$\alpha$ shows an unusual flat-topped emission profile.
A photospheric spectrum showing a P-Cygni H$\alpha$ line with flat-topped profile 
 is indicative that the H$\alpha$ emission mostly is produced in a detached region. 
Detached atmospheres are not common in SNe (\cite{jef90}) and this is the first 
time we see such a structure in a type II supernova. This could be caused by a
temporary peak in the density profile of the hydrogen layer. A flat-topped profile 
may be also produced  in a cool dense shell (CDS) that forms at the interface between the SN ejecta and 
the wind produced by the progenitor (\cite{chu07}). 

The evolution of SN 1999ga is slow and the two subsequent spectra show basically the same features as the first spectrum, plus much stronger nebular lines of
[O I] $\lambda$$\lambda$6300-6364 and the classical feature around 7300\AA~due to a blend of 
[Ca II] $\lambda\lambda$7291,7324, [O II] $\lambda\lambda$7320,7230 and [Fe II]. The H$\alpha$ feature
is narrower and, in contrast to the December 29 spectrum, has evolved developing a rounded profile. 
In the ejecta-wind interaction scenario of Chugai et al. (2007), this evolution of H$\alpha$ is expected to be accompained by 
an absorption component that progessively shifts toward redder wavelengths. This is not observed in SN 1999ga, making 
the CDS scenario quite unlikely.

Interestingly, the two intermediate spectra of SN 1999ga show H$\alpha$ with a slightly blue-shifted peak (see Fig. \ref{Fig4}). Blue-shifted emission peaks are observed 
in many young type II SNe (see e.g. \cite{tura93b}) and this in agreement with theoretical predictions (\cite{jef90}, see also \cite{luc07}),
but unexpected at late phases, unless caused by the formation of dust in the SN ejecta (like in SN 1999em, \cite{abou03}). However the dust formation in the SN
ejecta is usually accompanied by an increased slope of the optical light curves and, eventually, an IR excess. Interestingly, in the case of SN 1999ga
a deviation from the $^{56}$Co decay slope is clearly visible only at later phases (after 450-500 days from the explosion, see Fig. \ref{Fig3}).

A further spectrum was obtained about 13 months after the first one. Surprisingly there is no significant evidence for H$\alpha$ or other
H Balmer lines, while the emission feature at 7300\AA~ is still quite prominent. This is indeed the only spectral line unequivocally visible
in this spectrum. 

In Fig. \ref{Fig5} three spectra of SN 1999ga are compared with those of the type IIL SN 1990K (\cite{cap95b}) and the type IIP SN 2004et
(\cite{sahu07})\footnote{The spectra of the comparison objects have been downloaded from the Online Supernova Spectrum Archive SUSPECT {\it
(http://bruford.nhn.ou.edu/~suspect/index1.html)}.}. 
The three SNe appear to have a rather similar spectra at $\sim$120 and $\sim$200 days. The $\sim$120 days spectra of all SNe 
show a prominent H$\alpha$ line, although only in SN 1999ga H$\alpha$ has a flat-topped profile. 
Subsequent spectra of the three SNe (phase of about 200 days) show a rather normal, rounded H$\alpha$ profile
which is expected in non-interacting SNe II. 
As a remarkable difference with the H-rich type IIP SN 2004et, very late spectra of SNe 1999ga and 1990K obtained around 15 and 17.5 months (respectively) after their explosions surprisingly do not 
show any prominent H$\alpha$ line (Fig. \ref{Fig5}, bottom panel), but $-$ this is more clearly visible in SN 1990K $-$  only a broad, weak bump around
6300-6400\AA, which is mostly due to the [O I] $\lambda$6300-6364 doublet. The only (relatively) prominent feature is that at $\sim$7300A,
possibly a blend of several forbidden lines ([CaII], [O II], [Fe II]). The lack of prominent H$\alpha$ at very late times supports 
the scenario of a progenitor star with a small residual H envelope and, hence, a type IIL classification for SN 1999ga.

Apart from the peculiar H$\alpha$ line profile in the earliest spectrum, there is little evidence that
the SN is interacting with a CSM.  The lack of the flattening in the optical light curves
typical of SNe which are strongly interacting with a CSM is in good agreement with this scenario. As a further support,
no emission was detected at the position of SN~1999ga in 6 cm (5170 MHz) images analysed by Harnett et al. (2004),
neither in images obtained during the 1990s, before the SN explosion, or in a post-explosion image obtained on 2000 December 31
(i.e. more than one year after the explosion). Therefore, if the earliest optical spectrum may eventually suggest the presence of material
lost by the progenitor in pre-SN mass loss events, the subsequent SN evolution allows us to exclude major ejecta-CSM
interaction episodes.

\section{The birthplace of SN 1999ga} \label{hb}

   \begin{figure*}
   \centering
   \includegraphics[angle=0,width=18.5cm]{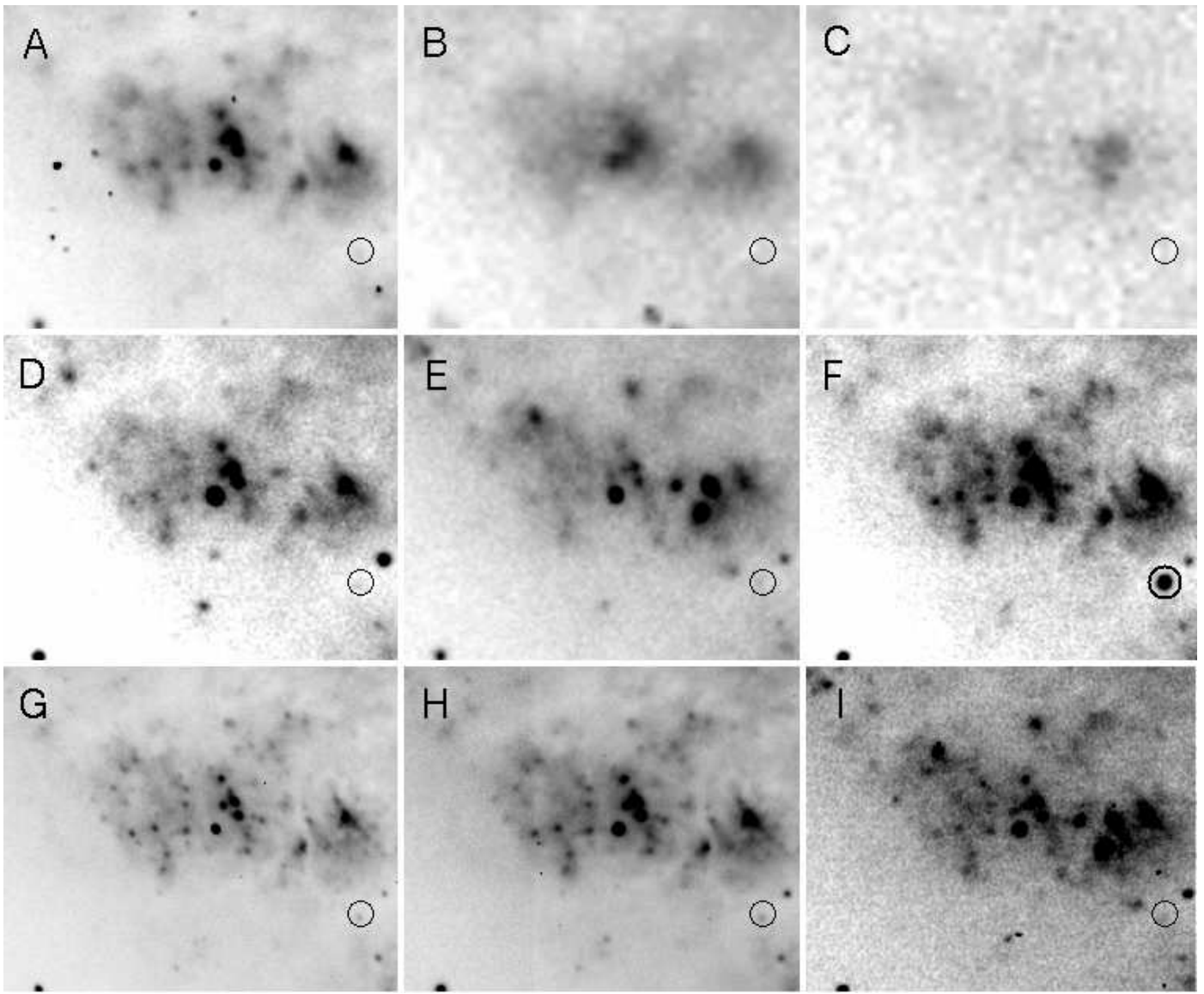}
   \caption{The explosion site of SN 1999ga: low-resolution frames obtained with different ground-based telescopes. (A): pre-explosion B band image obtained on March 1, 1995 with the 1.5m CTIO telescope. 
   (B) pre-explosion V band image obtained  March 4, 1990 with the 1m Telescope at Siding Spring Observatory (SSO).
   (C) pre-explosion H$\alpha$ image obtained on January 31, 1989  at the SSO 1m Telescope.
   (D) pre-explosion I band image obtained on  March 1, 1995 with the 1.5m CTIO telescope.
   (E) pre-explosion unfiltered image obtained on February 22, 1995 at the Anglo Australian Telescope.
   (F) B band image of SN 1999ga obtained on December 29, 1999 with the ESO/MPI 2.2m Telescope in La Silla.
   (G) post-explosion B band image obtained on January 30, 2006  with the ESO/MPI 2.2m Telescope. 
   (H) post-explosion V band image obtained on January 29, 2006  with the ESO/MPI 2.2m Telescope. 
   (I) post-explosion H$\alpha$ image obtained on January 28, 2006  with the ESO/MPI 2.2m Telescope. The SN location is marked by a circle.
   Only an extended source (denoted as {\it source A} in the text) is visible both in the pre-
    and post-explosion images in the SN vicinity. In all figures, north is up, east is to the left.}
              \label{Fig6}
    \end{figure*}

\begin{figure}
   \centering
   \includegraphics[angle=0,width=9cm]{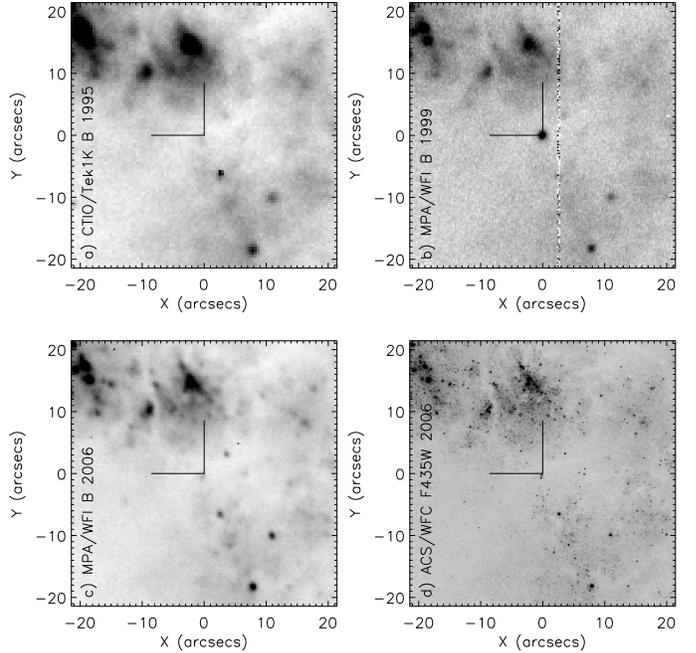}
   \caption{The explosion site of SN 1999ga in the B band. Top-left: low resolution, pre-explosion CTIO 1.5-m telescope image
   obtained on March 1, 1995. Top-right: low-resolution
   SN image obtained on December 29, 1999 with the 1.54-m Danish Telescope (La Silla, Chile) equipped with DFOSC.  
    Bottom-left: 
   late time low resolution frame obtained on January 30, 2006 with the 2.2m ESO/MPI Telescope (plus Wide Field Camera) in La
   Silla.    Bottom-right: late time (October 21, 2006) ACS HST image (filter F435W) of the explosion region. 
   Information on these images is reported in  Table \ref{Tab4}. All images are centered at the SN position
   and oriented such that north is up and east is to the left.}
              \label{Fig7}
    \end{figure}

A method to understand the nature of the star which exploded as SN 1999ga is to study the site of explosion, trying to derive 
information on the progenitor via the direct detection of the star (or constraining robust detection limits) in available pre-explosion images 
(e.g. \cite{sma04,van03,mau05,sma08}). 
The site of explosion of SN 1999ga was occasionally monitored in the past using small-size telescopes (see Tab. \ref{Tab4}).
A large sample of low-resolution archive images obtained using different filters and showing the explosion site over a period of about 17 years
is presented in Fig. \ref{Fig6}, while in Fig. \ref{Fig7} the region of the SN is shown in a sequence of B band 
images obtained before, during and after the SN outburst, including a high-resolution HST image of the post-explosion site (panel D).

In particular, pre-explosion images were obtained during the 1990s at the 1m Telescope and the 3.9m Anglo Australian Telescope
at Siding Spring Observatory (Australia), and the 1.5m Telescope of the Cerro Tololo Inter-American Observatory 
(CTIO) in Chile (Fig. \ref{Fig6}, inserts A to E, and Fig. \ref{Fig7}, insert A). Some of the images here analysed were published in \cite{ryd93}. 
The best-quality  pre-explosion images are those in the B and I bands obtained on March 1995 at the CTIO 1.5m-Telescope. 
Unfortunately, most pre-SN frames are not deep enough to provide robust detection limits for the putative progenitor star.

One problem is the SN location, which is on the northern edge of an  elongated luminous region (that will be labelled hereafter as {\it source A}, 
see Fig. \ref{Fig6} and Fig. \ref{Fig7}), visible in most pre- and post-explosion images. 
This region is particularly luminous in the B band, indicating that it is probably associated with luminous blue stars. 
Interestingly, {\it source A} is faint in the H$\alpha$ images.
This, together with the lack of evidence in the SN spectra of narrow lines from the galactic background, suggests
a stellar nature (possibly stellar clusters) for this vast source near the SN location, rather than a star-forming region. 
Since the SN exploded at the edge of {\it source A}, it is very difficult to extrapolate the flux 
contribution of a single star from that of the entire environment. We measured the integrated magnitudes of the unresolved
{\it source A} in the ground-based images before and several years after the SN explosion, and we found (within the errors)
no significant differences. This implies that the contribution of the progenitor of SN 1999ga to the flux of the whole
extended region was negligible.
 The magnitudes of {\it source A} as derived in the low resolution images are reported in Tab. \ref{Tab4} (column 6).

We also collected some deep post explosion images (Fig. \ref{Fig6}, inserts G, H, I; Fig.\ref{Fig7}, insert C) obtained 
with the ESO/MPI 2.2m Telescope in  January  (under the proposal
ID: 076.C-0888, PI: Y. Bialetski) and November 2006 (reserved MPI time,  program ID 078.A-9046(A), PI: W. Hillebrandt). We used the  best seeing B-band 
images (those of January 2006) as templates in our attempt to recover the progenitor star in the March 1995 B-band image.
After combining the best seeing images obtained at the same epoch, and after geometrical and photometric registration of the pre- and
post-explosion images, we subtracted the latter image from the former. With this procedure, the whole host galaxy and, hence, also 
the emitting region in the SN vicinity were removed. 
Nevertheless, any attempt to recover the progenitor star in the B-band failed, since there was no evidence of the progenitor of SN 1999ga  
in the subtracted image at the SN position at a B-band magnitude brighter than 23.15.
Similar attempts were made with other images, and none showed evidence of the presence of any star 
at the position of the SN. Adopting the distance and reddening estimates discussed in Sect. \ref{ga}, we 
obtain M$_B \gtrsim$ -9.3 which is not, obviously, a very significant detection limit.

\begin{figure*}
   \centering
   \includegraphics[angle=0,width=18cm]{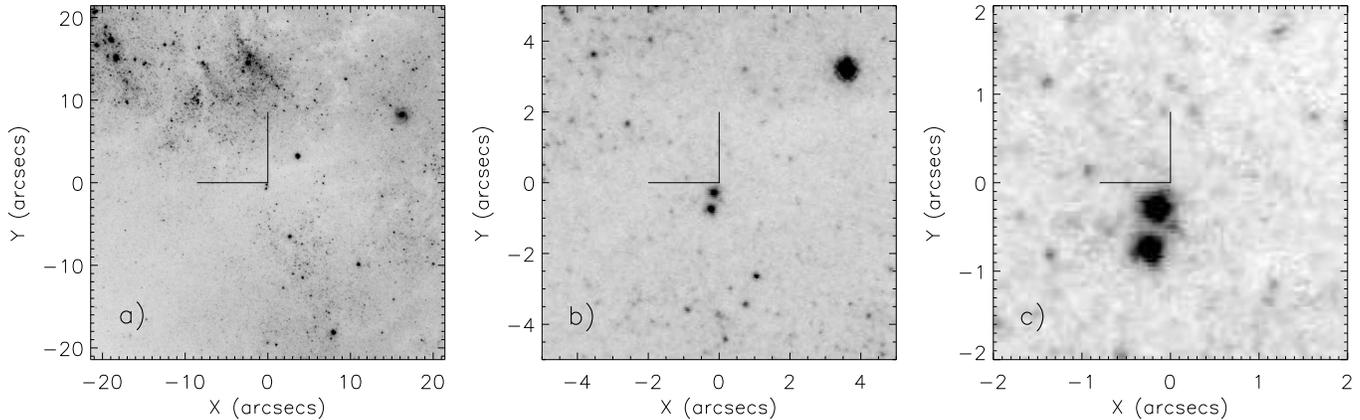}
   \caption{HST F814W band (roughly I) image obtained 7 years after the SN explosion (left panel) and blow ups of the SN region: $\times$4 (central panel) and
   $\times$10 (right panel). The SN position is marked by the cross-hairs, and no source is visible at that position. The high resolution of the HST/ACS images allows
   us to resolve {\it source A} into two
   main components (probably two stellar clusters) plus other (though much fainter) sources. 
   The images are centered at the SN position. North is up, east to the left.}
              \label{Fig8}
    \end{figure*}

In order to study in more detail the structure of the extended {\it source A}, 
 the site was targeted by the Hubble Space Telescope (HST) on October 21st, 
2006, 7 years after the SN explosion (proposal ID: 10803, PI: S. J. Smartt). {\it Source A}, which appears to be elongated (roughly) in
the North-South direction in the ground based images, in reality consists of two major sources 
(plus a few much fainter sources visible in their vicinity),  as one can clearly gather from the 
high-resolution HST images (see blow-up panels of the F814W HST/ACS image, Fig. \ref{Fig8}).
Both these sources are extended with FWHM  which is about twice that of the stellar PSF, so that
we suggest that they are both compact stellar clusters. 

We performed PSF-fitting photometry using the {\it DOLPHOT}\footnote{{\it DOLPHOT} is a stellar photometry package that was adapted from HSTphot 
(\cite{dol00}) for general use.} package.  {\it DOLPHOT} classifies the southern source as a single, point-like object (type 1) and the northern 
source as extended (type 5).  However, rather significant residuals are visible at the sites of both objects after PSF subtraction, and we conclude 
that no one of them is well fit by this photometry package.

Aperture photometry of these objects is complicated by the non-stellar nature of their PSFs.  This makes it difficult to define aperture corrections 
and the objects are too close one another to use a large aperture.

The {\it BAOLAB} task {\it ISHAPE} (\cite{ssl99,ssl04}) 
was used to determine the intrinsic size and shape of these assumed clusters.  
{\it ISHAPE} convolves an analytical function that is assumed to represent the actual source 
(e.g. a delta function for a star) with the stellar PSF.  {\it ISHAPE} then compares this convolved function to the data to
determine the best fit and therefore the intrinsic FWHM of the source function.
An elliptical Moffat function with a power index of 1.5 was used to fit our proposed clusters.  
The resulting parameters for the southern and northern sources (computed for all HST images) are reported in Tab. \ref{Tab5}.

\begin{table}
\caption{Basic parameters for the two putative clusters forming {\it Source A}, as derived
by BAOLAB/ISHAPE in each of the three HST filters.}\label{Tab5}
\centering
\begin{tabular}{cccc} \hline \hline
Source & F435W & F658N & F814W \\ \hline
\multicolumn{4}{c}{FWHM of semi-major axis (pixels)}\\ 
Northern  &     1.94   &      1.96   &      2.22 \\
Southern  &     1.16   &      1.70   &      1.37 \\

\hline

\multicolumn{4}{c}{Ratio of major/minor axis}\\
Northern  &     0.98   &      0.61   &      0.95 \\
Southern  &     0.72   &      0.44   &      0.83 \\

\hline

\multicolumn{4}{c}{Position Angle (degrees)}\\ 
Northern   &    -43.9   &     -2.6     &    -48.1 \\
Southern   &    -14.1   &     -69.0    &    -14.6 \\ 

\hline

\multicolumn{4}{c}{Photometry (Vegamag)}\\
Northern  &     23.49   &     22.35    &    21.95 \\
Southern  &     23.62   &     22.46    &    22.09 \\ 
\hline\hline
\end{tabular}
\end{table}

It is worth noting that the intrinsic size, major/minor axis ratio and position angle of
each object are consistent between the two broadband filters, while in H$\alpha$
(F658N) they appear to have completely different shapes and orientations.
In the case of the southern source, its extent in the H$\alpha$ image is 
significantly larger than those derived in the broadband images.  These measurements suggest
that there is nebular H$\alpha$ emission which is significantly elongated
relative to the the cluster.

The {\it BAOLAB} task {\it MKCMPPSF} was used to calculate synthetic PSF matching
of these two sources in each filter.  These were added to fake images and
used to calculate appropriate aperture corrections for the objects.  This
resulted in the corrected aperture photometry reported in Tab. \ref{Tab5} (bottom).

With the distance and extinction toward SN 1999ga adopted in this paper (see Sect. \ref{ga}), 
we obtain absolute magnitudes of F435W $\sim$ -8.9 and F814W $\sim$ -9.5 for 
both objects\footnote{Note that the adopted extinction is not necessarily
appropriate for these proposed clusters. However, without photometry in more
filters and/or spectra of these objects, it is not possible to provide a direct
measurement of the extinction towards the two sources.}, 
which are certainly brighter than we would expect single stars to be.


Finally, the half-light/effective radii of the clusters from the F814W
image was computed and found to be
R(eff)$_{Northern}$ = 9.5 $\pm$ 0.4 pc and
R(eff)$_{Southern}$ = 5.5 $\pm$ 0.3 pc.
The northern cluster has a projected separation from the SN position of
about 28 $\pm$ 5 pc (7.3 $\pm$ 1.3 pixels). Although previous studies allowed 
constraints to be placed on SN progenitors by estimating the ages and the
main sequence turn-off masses of host stellar clusters (\cite{mai04,cro08a,cro08b,ost08}),
in the case of SN 1999ga the progenitor was probably too far 
from the northern cluster to claim that it was coeval.
Therefore, despite the large amount of pre- and post-explosion images
available for NGC 2442, no robust information can be derived for the progenitor
star of SN 1999ga, since  the only thing we can exclude is that the precursor was a very luminous star (e.g. a luminous
blue variable).

\section{Discussion and Conclusions} \label{dc}

Despite lacking early-time photometric monitoring of SN 1999ga, which would have allowed us to conclusively discriminate between 
the different subtypes of SNe II, we believe that spectroscopy has provided enough evidence to support a designation of type IIL for this object.
The  non-detection of H$\alpha$ in the latest spectrum, in particular, 
indicates that the residual H envelope of the exploding star 
was not very massive (although probably more massive than that of a type IIb SN). 

A sub-division of SNe IIL into two groups (bright, like SN 1979C, and regular events, depending on their 
intrinsic luminosities) was proposed by a number of authors (\cite{you89,gas92,pat94}).
Richardson et al. (2002) estimated the average absolute magnitudes for bright and normal SNe IIL to be around M$_B \approx$ -19.3 and M$_B \approx$ -17.6,
respectively. It is evident from a simple check of the absolute magnitudes of SN 1999ga at any time that this object is much fainter than regular SNe IIL.
This may possibly indicate the existence of a low-luminosity tail in the luminosity distribution of type IIL SNe, similar to that
already observed in SNe IIP (\cite{ham03,pasto04}). Even if the peak luminosity can be questioned because of the late discovery of SN 1999ga,
the radioactive tail is much fainter than that observed in other SNe IIL (see Fig. \ref{Fig3}) and is comparable in luminosity with those of 
low-luminosity SNe IIP (\cite{pasto04}).
This faint radioactive tail is consistent with the ejection of a small mass of $^{56}$Ni ($\sim$10$^{-2}$ M$_\odot$), an amount which is only marginally 
higher than that reported for low-luminosity SNe IIP (\cite{pasto04,pasto09}).
However, in stark contrast to faint SNe IIP, the broad spectral features observed in the SN 1999ga spectra
are indicative of high-velocity ejecta (5000-6000 km s$^{-1}$, see Sec. \ref{sp}). These two observed quantities suggest a 
normal explosion energy and moderate mass of the ejected material (including radioactive $^{56}$Ni) for SN 1999ga.
There is indeed robust evidence from spectroscopy that the mass of the H envelope was rather small and/or that there
was a non-negligible amount of CSM, resulting from mass loss episodes during the late stages of the stellar life.
Several attempts have been made to estimate the explosion parameters of past SNe IIL. 
The observed evolution of the nebular H$\alpha$ line in the cases of the bright SNe 1980K and 1990K (\cite{cap95b} and references therein) 
was well reproduced by the models of \cite{chu91} with 5M$_\odot$ of ejecta, canonical explosion energy (E$_0$ = 10$^{51}$ erg) and intermediate
$^{56}$Ni mass (M($^{56}$Ni) = 0.075M$_\odot$). Slightly higher values of the above parameters are probably necessary to account for the higher luminosity of H$\alpha$
in SN 1979C, even though the luminous light curve peak probably did not result from anomalous explosion parameters, but through reprocessing of UV 
light in a shell generated by pre-SN wind (\cite{bar92}).

   \begin{figure}
   \centering
   \includegraphics[angle=0,width=9.0cm]{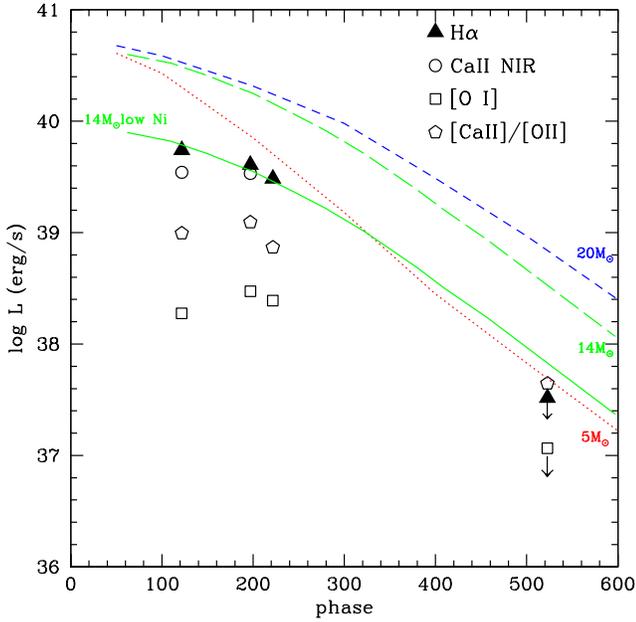}
   \caption{Luminosity evolution of the main nebular features in the spectra of SN 1999ga. Models of \protect\cite{chu91} showing 
   the evolution of H$\alpha$ in SNe II under the assumption of pure radioactive decay of $^{56}$Co are also shown. The dotted red line
   marks the H$\alpha$ luminosity evolution for an ejected mass of 5M$_\odot$, the long-dashed green line for 14M$_\odot$ and the dashed blue line for 20M$_\odot$. 
   All models were computed with an explosion energy of 10$^{51}$ erg and an ejected $^{56}$Ni mass of 0.075M$_\odot$.
   The solid green lines is the 14M$_\odot$ model, but rescaled to 0.015M$_\odot$ of $^{56}$Ni.}
              \label{Fig9}
    \end{figure}

In Fig. \ref{Fig9} we compare the luminosity evolution of the H$\alpha$ line in SN 1999ga with models of H$\alpha$ luminosity 
expected in type II SNe, assuming that the source of the luminosity is purely radioactive decay (Chugai 1990, 1991).
The models were obtained adopting ejected masses of 20M$_\odot$ (blue dashed line), 14M$_\odot$ (green long-dashed line), 
5M$_\odot$ (dotted red line), and computed with E$_0$ = 10$^{51}$ erg and M($^{56}$Ni) = 0.075M$_\odot$.
In Fig. \ref{Fig9} the luminosity evolution of other typical nebular lines ([O I]$\lambda$$\lambda$6300-6364, the [Ca II] plus [O II] blend
around 7300 \AA~and the Ca II NIR triplet) is also shown.
We note that the H$\alpha$ luminosities are systematically lower than those predicted by Chugai's models with 0.075M$_\odot$ of $^{56}$Ni,
although the early decline rate is quite consistent with that of the 14M$_\odot$ model. 
We therefore rescaled the 14M$_\odot$ model of Chugai (1990) to M($^{56}$Ni) = 0.015M$_\odot$. 
This value is consistent with the $^{56}$Ni mass deduced from the late time light curve of SN 1999ga (see Sect. \ref{lc}). 
The rescaled 14M$_\odot$ model is shown in Fig. \ref{Fig9} 
(solid green line), and fits reasonably well to the observed H$\alpha$ luminosities of SN 1999ga only at earlier epochs, 
while it fails to match the late-time H$\alpha$ detection limit.
A lower ejected mass would help
to better reproduce the faster H$\alpha$ luminosity decline observed at late time, although it would fail to fit the earliest
observed point of SN 1999ga. A reasonable range for the total ejected mass of SN 1999ga is therefore 6-8M$_\odot$, with
only 1-2$_\odot$ of H.

SNe IIL belong to a sequence of supernova types produced by progenitors with increasing mass loss occurring during the late phases of their evolution,
i.e. SN IIP $\rightarrow$ SN IIL $\rightarrow$ SN IIb $\rightarrow$ SN Ib $\rightarrow$ SN Ic (e.g. \cite{nom97,che06}).
However, the nature of their progenitor stars is still unclear. Two channels have been proposed for type IIL SNe: single, massive progenitors 
(M$_{ZAMS} \geq$ 20M$_\odot$), and  lower mass stars  (7-10M$_\odot$, \cite{swa91}) in binary systems, where the mass loss is triggered by the companion 
star.

Unfortunately, the lack of high-resolution, deep, pre-explosion HST images prevents us from providing robust constraints on the nature of the progenitor
of SN 1999ga. With the remarkable exception of the detection of the K supergiant progenitor of SN 1993J (\cite{ald94,mau04,mau09}), 
most attempts to identify the progenitors of  stripped (or partially stripped) envelope core-collapse
SNe have failed (\cite{sma02,van03,mau05,mau05b,gal05,cro07,cro08a}) or are disputed (\cite{gal07,cro08b}).
However, the lack of any detection in ground based pre-explosion images, the analysis of the surrounding
environment, and the observed SN properties (unusual spectral and H$\alpha$ luminosity evolutions, sub-luminous light curves, small $^{56}$Ni mass) all suggest 
that the precursor of SN 1999ga at the time of core-collapse was likely a moderate-mass star (8-10M$_\odot$), with a moderate-to-low mass residual H envelope.
The fact that the H lines and not the He lines dominate the SN spectra in the early nebular phase is indicative
that the star retained a significant mass (1-2M$_\odot$) of H at the time of the explosion. This is more than the few $\times$ 10$^{-1}$M$_\odot$
expected for type IIb SNe (e.g. SN 2008ax, see \cite{pasto08} and references therein), but surely less than the several solar masses
estimated for a normal type IIP SN (e.g. \cite{nad03}). Although the ejecta mass implies a moderate mass star (8-10M$_\odot$) at
explosion, the initial mass is somewhat uncertain. As the hydrogen mass is low, the progenitor likely lost mass through either stellar
winds or mass-transfer to a binary. In either case, the amount of mass-loss could be significant (e.g. 5-10M$_\odot$). The pre-explosion
images are not deep enough to distinguish between these two scenarios and the SN is not close enough to the compact clusters
to assume that it is coeval with their stellar population.

SN 1999ga is probably the first (relatively) under-luminous, $^{56}$Ni poor core-collapse supernova that can be 
classified as type IIL. The velocity of the ejecta, as constrained by the width of the spectral lines, is not as
low as observed in faint type II-P SNe (\cite{pasto04}). This provides further evidence, in contrast with SNe IIP,
for a moderate mass ejected by SN 1999ga.
The detection of this kind of objects is a rare event, probably because of their faint nature
coupled with a fast photometric evolution. The discovery of SN 1999ga
should be seen as motivation for searches for more extreme, fast-evolving, sub-luminous, envelope-stripped core-collapse SNe. These are expected 
to occur (e.g. \cite{woo93}) but, apart from a few remarkable exceptions (see \cite{val09}), usually elude detection.
Similar SNe have been also proposed to be responsible for a class of long $\gamma$-ray bursts that do not show 
evidence of associated SNe (\cite{mdv06,fyn06,gal06,geh06,ofe07,dado08}).

\begin{acknowledgements}

We ackwnoledge the anonymous referee for useful comments on the manuscript.
This paper is based on observations made with the NASA/ESA Hubble Space Telescope,
with the Perth-Lowell 0.61m Cassegrain Telescope of  the Perth Observatory (Australia), 
and at the 1.54m Danish Telescope, the ESO/MPI 2.2m Telescope and the 3.6m Telescope at 
the European Southern Observatory (La Silla, Chile) under program IDs 
65.H-0292 (PI: M. Turatto)
66.D-0683 (PI: M. Turatto).
Support for program
GO-10803 was provided by NASA through a grant from the Space Telescope Science Institute,
which is operated by the Association of Universities for Research in Astronomy, Inc., under
NASA contract NAS 5-26555.
This paper is also partly based on observations made with the European Southern Observatory
telescopes obtained from the ESO/ST-ECF Science Archive Facility. We also made
use of data obtained from the Anglo Australian Telescope Archive which is 
maintained as part of the CASU Astronomical Data Centre at the Institute of 
Astronomy, Cambridge.
SB, EC and MT are supported by the Italian Ministry of Education via the PRIN 2006 n.2006022731-002.
SM acknowledges funding from the Academy of Finland (project 8120503).

\end{acknowledgements}

\end{document}